\documentclass[twocolumn]{aastex631}
\shorttitle{ACA [CI] observations in NGC~7679}
\shortauthors{Michiyama et al.}
\graphicspath{{./}{figures/}}
\usepackage{amsmath}
\usepackage{comment}

\begin{document}

\title{From Non-Detection to Detection: Atacama Compact Array Mosaic Observations of Faint Extended [\ion{C}{1}] Emission in NGC~7679}

\correspondingauthor{Tomonari Michiyama}
\email{t.michiyama.astr@gmail.com}

\author[0000-0003-2475-7983]{Tomonari Michiyama}
\affiliation{Faculty of Information Science, Shunan University, 843-4-2 Gakuendai, Shunan, Yamaguchi, 745-8566, Japan}

\author{Toshiki Saito}
\affiliation{Faculty of Global Interdisciplinary Science and Innovation, Shizuoka University, 836 Ohya, Suruga-ku, Shizuoka 422-8529, Japan}

\author{Kouichiro Nakanishi}
\affiliation{National Astronomical Observatory of Japan, National Institutes of Natural Sciences, 2-21-1 Osawa, Mitaka, Tokyo, 181-8588, Japan}
\affiliation{Department of Astronomical Science, The Graduate University for Advanced Studies, SOKENDAI, 2-21-1 Osawa, Mitaka, Tokyo, 181-8588, Japan}

\author{Daisuke Iono}
\affiliation{National Astronomical Observatory of Japan, National Institutes of Natural Sciences, 2-21-1 Osawa, Mitaka, Tokyo, 181-8588, Japan}
\affiliation{Department of Astronomical Science, The Graduate University for Advanced Studies, SOKENDAI, 2-21-1 Osawa, Mitaka, Tokyo, 181-8588, Japan}

\author{Ken-ichi Tadaki}
\affiliation{Faculty of Engineering, Hokkai-Gakuen University, Toyohira-ku, Sapporo 062-8605, Japan}

\author{Juan Molina}
\affiliation{Instituto de Física y Astronomía, Universidad de Valparaíso, Avda. Gran Bretaña 1111, Valparaíso, Chile}

\author{Bumhyun Lee}
\affiliation{Department of Astronomy, Yonsei University, 50 Yonsei-ro, Seodaemun-gu, Seoul 03722, Republic of Korea}
\affiliation{Korea Astronomy and Space Science Institute, 776 Daedeokdae-ro, Daejeon 34055, Republic of Korea}

\author{Ming-Yang Zhuang}
\affiliation{Astronomy Department, University of Illinois Urbana-Champaign, Urbana, IL 61801, USA}

\author{Junko Ueda}
\affiliation{National Astronomical Observatory of Japan, National Institutes of Natural Sciences, 2-21-1 Osawa, Mitaka, Tokyo, 181-8588, Japan}

\author{Takuma Izumi}
\affiliation{National Astronomical Observatory of Japan, National Institutes of Natural Sciences, 2-21-1 Osawa, Mitaka, Tokyo, 181-8588, Japan}
\affiliation{Department of Astronomical Science, The Graduate University for Advanced Studies, SOKENDAI, 2-21-1 Osawa, Mitaka, Tokyo, 181-8588, Japan}

\author{Luis C. Ho}
\affiliation{Kavli Institute for Astronomy and Astrophysics, Peking University, 5 Yiheyuan Road, Haidian District, Beijing 100871, P.R.China}
\affiliation{Department of Astronomy, School of Physics, Peking University, Beijing 100871, China}








\begin{abstract}
We report the detection of [\ion{C}{1}]~$^3P_1$--$^3P_0$ emission in the nearby galaxy NGC~7679 using the Atacama Compact Array (ACA) of the Atacama Large Millimeter/submillimeter Array (ALMA). In \citet{Michiyama2021ApJS..257...28M}, [\ion{C}{1}]~$^3P_1$--$^3P_0$ emission in NGC~7679 was reported as undetected based on ACA observations conducted in 2019 (ALMA Cycle~6). These observations had $\sim1$ minute on-source time and used a single pointing with a field of view (FoV) of $\sim20\arcsec$. In 2023 (Cycle~9), we carried out mosaic observations using seven pointings with an FoV of $\sim27\arcsec$ and 4--5 minutes on-source per pointing. The additional data have significantly improved the line sensitivity, $uv$-sampling, and noise uniformity across the galaxy disk. Our Cycle~9 observations confirm the presence of extended [\ion{C}{1}]$^3P_1$--$^3P_0$ emissions in NGC 7679, which was completely missed in the Cycle~6 observations due to insufficient sensitivity and $uv$-sampling. This highlights the basic technical challenges of estimating the total flux by interferometric observations with sparse $uv$ sampling.

\end{abstract}

\keywords{Extragalactic astronomy (506) --- Submillimeter astronomy (1647) --- Interstellar medium (847) }


\section{Introduction}
The lower forbidden $^3P$ fine structure transition of atomic carbon, denoted as [\ion{C}{1}]~$^3P_1$--$^3P_0$ or [\ion{C}{1}]~(1--0), emits at a rest frequency of 492~GHz ($\lambda=609\,{\rm \mu m}$). 
\citet{Michiyama2021ApJS..257...28M} conducted a survey project observing the [\ion{C}{1}]~(1--0) and CO~(4--3) emission in nearby ultra/luminous infrared galaxies (U/LIRGs) during the Atacama Large Millimeter/submillimeter Array (ALMA) Cycle~6 period using the 7-m array of the Atacama Compact Array (ACA), without the total power (TP) array.
This effort identified three galaxies: NGC~6052 \citep{Michiyama2020ApJ...897L..19M}, NGC~7679, and ESO~467-G027, showing notably faint [\ion{C}{1}]~(1--0) emissions compared to that of CO~(4--3).
In this report, we present the detection of [\ion{C}{1}]~(1--0) toward NGC~7679 and briefly discuss the difficulties we faced.

\section{Data}
\subsection{Observations}
The target, NGC~7679, is a nearby galaxy with a redshift of \(z=0.0171\). 
The observing frequency for redshifted [\ion{C}{1}]~(1--0) is 483.887\,GHz.
The ACA observations targeting [\ion{C}{1}]~(1--0) emission in NGC~7679 were performed during the Cycle~9 season (ALMA project ID: 2022.1.00746.S; PI: T. Michiyama).
On June 15, 2023, mosaic observations (seven pointings) using the 7-m array were conducted, with an on-source time of about 4--5 minutes per pointing. This on-source time is longer than that in the previous Cycle 6 observation, which was $\sim1$ minute.
The TP array observations were conducted on October 13, 2022, and April 13, 2023.

\subsection{Data Reduction}
The calibrated visibility data were obtained by running the standard pipeline calibration scripts provided by ALMA using {\tt CASA} \citep{CASA2022PASP..134k4501C}. The imaging processes were conducted using {\tt CASA} 6.5.0-15, with a velocity resolution of 30\,km\,s$^{-1}$, a cell size of 0\farcs25, an imaging size of 256 pixels, and Briggs weighting with a robust parameter of 0.5. The clean masks for each channel were manually determined. We performed the same clean task in both Cycle~6 (single-point) with a FoV of $\sim20\arcsec$ and Cycle~9 (7-pointing mosaic) with a FoV of $\sim27\arcsec$. For a fair comparison, we used the same maximum and minimum baselines ([14, 69]\,k$\lambda$) in both datasets (see the top panel of Figure~\ref{fig:Cy9Cy6}), and the synthesized beam was smoothed to 3\farcs5.
The rms noise level of the data cube was 
$\sim0.02$\,Jy\,beam$^{-1}$ in the Cycle~9 data and $\sim0.08$\,Jy\,beam$^{-1}$ in the Cycle~6 data.
The moment~0 map was made by integrating the channels of [$-150$, $150$]\,km\,s$^{-1}$ without applying any masking.

\subsection{Results}
The velocity integrated intensity (moment~0) map of the Cycle~9 data clearly shows the detection of the [\ion{C}{1}]~(1--0) emission line (see the middle panel of Figure~\ref{fig:Cy9Cy6}).
We measure an integrated flux of $123\pm10$\,Jy\,km\,s$^{-1}$ by applying a two-dimensional Gaussian fit to sources in the moment~0 image using the {\tt imfit} task in {\tt CASA}.
It is important to note that this integrated flux might not represent the ``total" [\ion{C}{1}]~(1--0), as the line flux of the tentatively detected [CI] by the TP array (Michiyama et al. in prep.) may be further higher than what we measured with the ACA here.

\section{Discussion}
\begin{figure*}[ht!]
\epsscale{1.1}
\plotone{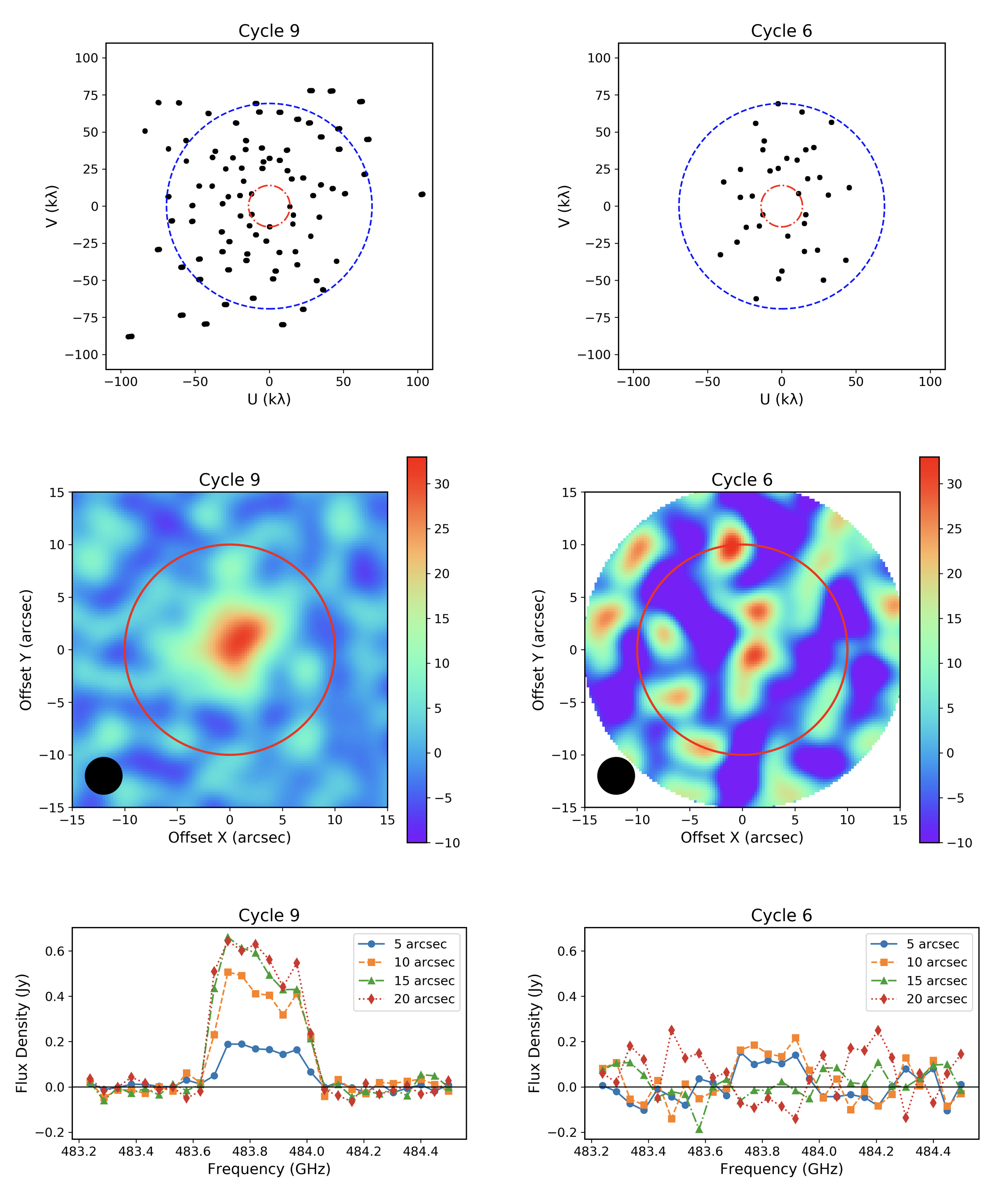}
\caption{The results of ACA observation targeting [\ion{C}{1}]~(1--0) emission in NGC~7679. The comparison between Cycle~9 (left) and Cycle~6 (right).  (Top:) The visibility distribution on the $uv$-plane. The blue and red dots represent the maximum and minimum $uv$ distances corresponding to baseline lengths in Cycle~6. (Middle:) The moment 0 map with the color scale of [-10, 33]\,Jy\,km\,s$^{-1}$\,beam$^{-1}$. The red line indicates a circle with a diameter of 20 arcseconds.
The black-filled circle indicates the synthesized beam of 3\farcs5.
(Bottom:) The lines indicate the spectra obtained with different apertures of 5 (blue), 10 (orange), 15 (green), and 20 (red) arcseconds. The black line represents the zero level. \label{fig:Cy9Cy6}}
\end{figure*}

Because the sensitivity in the Cycle~9 observations is four times better than that in Cycle~6, the new detection is reasonable.
When we create an image using a single scan for the central pointing in the Cycle~9 dataset, we detect no emission consistent with the Cycle~6 observation.
The bottom panel of Figure~\ref{fig:Cy9Cy6} highlights that the source remains undetected, even when simply summing the values of each pixel in the channel map.
For instance, in the case of a smaller aperture ($5\arcsec$ circle shown by the blue line in the bottom panel of Figure~\ref{fig:Cy9Cy6}), the spectrum in the Cycle~6 data appears similar to that in the Cycle~9 data around the line detection frequency range. However, when integrating the spectrum over the pixels within a larger aperture, the Cycle~9 spectrum shape is not reproduced.

A concern arises from the contradiction between the upper limit derived from the Cycle~6 moment-0 map and the integrated flux calculated from the Cycle~9 data.
This is due to a fundamental weakness when dealing with spatially extended sources.
As shown in equation~(4) of \citet{Michiyama2021ApJS..257...28M}, an emission line flux upper limit can be estimated as:
\begin{equation}\label{eq}
S_{\rm line}\Delta v < 3\sigma_{\rm ch}\Delta V_{\rm ch}\sqrt{\frac{N_{\rm ch}N_{\rm S}}{N_{\rm B}}},
\end{equation}
where $\sigma_{\rm ch}$ is the RMS noise level in the channel map, $\Delta V_{\rm ch}$ is the velocity resolution, $N_{\rm ch}=\mathrm{FWHM}/\Delta V_{\rm ch}$ is the number of channels within the FWHM, $N_{\rm S}$ is the number of pixels corresponding to the source size, and $N_{\rm B}$ is the number of pixels within the synthesized beam. If the source distribution is unknown, assumptions must be made regarding the line profile and source size. For example, assuming $\mathrm{FWHM}=150$\,km\,s$^{-1}$ and $N_{\rm S}/N_{\rm B}=3$, the $3\sigma$ upper limit is calculated as $S_{\rm line}\Delta v < 30$~Jy\,km\,s$^{-1}$ for $\Delta V_{\rm ch}=30$\,km\,s$^{-1}$ and $\sigma_{\rm ch}=0.08$~Jy\,beam$^{-1}$, based on our Cycle~6 data. This upper limit is inconsistent with the integrated flux measured in Cycle~9 data.
However, the equation~(\ref{eq}) assumes that the [\ion{C}{1}]~$^3P_1$--$^3P_0$ emission exhibits a spatial distribution detectable within the limited $uv$-sampling and FoV of a single-pointing observation, as in Cycle~6.
If the emission is spatially extended, it is not surprising that the flux detected in Cycle~9 was completely missed in Cycle~6. The new detection can be attributed to the longer integration time and mosaicking in Cycle~9, which improved the $uv$-coverage and ensured uniform noise across the entire map (Figure~\ref{fig:Cy9Cy6}).

As CO~(4--3) observations were not conducted in Cycle~9, the new detection of [\ion{C}{1}](1--0) does not fully negate the conclusions of a low [\ion{C}{1}]~(1--0)/CO~(4--3) ratio reported by \citet{Michiyama2021ApJS..257...28M}. This is because CO~(4--3) observations from Cycle 6 may have also missed the diffuse components.
Finally, we note that \citet{Michiyama2020ApJ...897L..19M} reported a non-detection of the [\ion{C}{1}]~(1--0) line in the nearby galaxy NGC~6052, which was observed in the same survey project as NGC~7679 in Cycle~6. Since the new ACA observations of NGC~6052 assigned in Cycle~9 were not completed, we cannot determine whether the faintness of the [\ion{C}{1}]~(1--0) line in NGC~6052 is real or not at this stage.


\renewcommand{\internallinenumbers}{}
\begin{acknowledgments}
\nolinenumbers
We thank Dr. Ran Wang for her valuable discussions.
T.M. was supported by a University Research Support Grant from the National Astronomical Observatory of Japan (NAOJ).
T.S. was supported by the Daiichi-Sankyo ``Habataku" Support Program for the Next Generation of Researchers.
K.N. acknowledges support from the JSPS KAKENHI grant No. 19K03937. 
This report makes use of the following ALMA data: ADS/JAO.ALMA 2022.1.00746.S and 2018.1.00994.S.

%
\vspace{5mm}
\facilities{ALMA}
\software{
ALMA Calibration Pipeline and CASA
}
\end{acknowledgments}

%




\bibliography{sample631}{}
\bibliographystyle{aasjournal}

\end{document}